# Electronic Commerce, Consumer Search and Retailing Cost Reduction

**Cristina Mazón**[1]

Universidad Complutense de Madrid, Facultad de Ciencias Económicas y Empresariales, Departamento de Fundamentos del Análisis Económico II, Campus de Somosaguas, 28223 Madrid, Spain

e-mail: cmazon@ccee.ucm.es; phone: 91-394-2918; fax 91-394-2591

**Pedro Pereira**

Departamento de Economía, Universidad Carlos III de Madrid, C./ Madrid, 126, 28903 Getafe, Madrid, Spain; e-mail: jppereira@jazzfree.com; phone: 91-544-1026; fax: 91-394-2613

**June 2001**

**Abstract:** This paper explains four things in a unified way. First, how e-commerce can generate price equilibria where physical shops either compete with virtual shops for consumers with Internet access, or alternatively, sell only to consumers with no Internet access. Second, how these price equilibria might involve price dispersion on-line. Third, why prices may be higher on-line. Fourth, why established firms can, but need not, be more reluctant than newly created firm to adopt e-commerce. For this purpose we develop a model where e-commerce reduces consumers' search costs, involves trade-offs for consumers, and reduces retailing costs.

**Key Words:** *Electronic-Commerce, Search, Cost Reduction, Retailing*

**JEL Classification:** *D43, D83, L13, L81, M31, O31, O33*

[1] We thank P.P. Barros, A. Cabrales, L. Corchón, R. Gatti, V. Karamychev, A. Kirman, J.L. Moraga, L. Moreno, and specially J. Sobel for useful comments, and E. Wasmer for giving us the *Yahoo Internet Life* example. The paper was presented at the Tinbergen Institute/Eramus University, I.S.E.G., U.A.B., C.E.M.F.I., W.Z.B., U.J.I, U.P.F., XV Jornadas de Economia Industrial 1999, Stony Brook 2000, E.E.A. 2000, E.A.R.I.E. 2000, A.S.S.E.T. 2000, and "The Economics of The Software and Internet Industries" Toulouse 2001. We thank Universidad Complutense, "Proyecto de Investigacion Complutense, # PR52/00-8892", for financial support.

# 1 Introduction

The Internet allowed the creation of a new retailing technology: ***electronic commerce*** (e-commerce)[2]. E-commerce has similarities with catalogue retailing. Without a physical shop, it offers products that cannot be physically inspected or immediately delivered, and are paid for usually with credit card. But e-commerce also has unique attributes. The Internet allows to cheaply store, search, and disseminate information; is available anywhere, anytime, for anyone who can accede to it; allows interactivity; provides perceptual experiences superior to those of a catalogue, but inferior to those of physical inspection; and serves as a transactions and physical distribution medium for information goods[3]. Due to this last aspect, conceivably, it will be in markets for information goods that e-commerce will have a bigger impact, as the recent evolution of markets for stocks, mortgages, or life insurance suggests (Bakos et al. (2000), Brown & Goolsbee (2000)).

Four facts have emerged from e-commerce's short history. First, typically, newly created, purely virtual firms, adopted e-commerce before established firms. In the book retail industry, new firms like Amazon adopted e-commerce before old firms like Barnes & Noble, and in the stockbroking retail industry, new firms like E*trade also adopted e-commerce before old firms like Charles Schwab. Are established firms intrinsically reluctant to adopt e-commerce? Second, physical shops responded to e-commerce, sometimes by lowering their prices to compete with virtual shops for consumers that buy on-line; other times they did not lower their prices, and concentrated on selling only to consumers with Internet access. Charles Schwab lowered the off-line fee from $65 to $30 to match the on-line fee (*New York Times*, August 16, 1999), which suggests the first price regime. And in 1999, Barnes & Noble and Borders matched within hours an Amazon 50% discount on best sellers on their virtual, but not on their physical shops (*The New York Times*, May 18, 1999). This suggests that established firms competed for consumers with Internet access on their virtual shops, but not on their physical shops, i.e., the second price regime. What explains these price regimes? Third, there is price dispersion on-line. Brynjolfsson & Smith (1999) find that established firms' virtual shops charge 8.7% more than new firms' virtual shops, and that on-line the price range is 33% of the average price[4]. If supposedly, the Web gives consumers access to perfect information, what explains price dispersion on-line? Fourth, prices are typically lower on-line. Brynjolfsson & Smith (1999) find that prices average 9-16% less on virtual shops than on physical shops. Are lower prices intrinsic to e-commerce?

---

[2] Transacting products based on the processing and transmission of digitized data over the network of computers that use the transmission control protocol/Internet protocol, TCP/IP.
[3] Goods that can be digitized, i.e., expressed as zeros and ones.

We believe these four facts are related, and develop a static, homogeneous product, partial equilibrium search model, that explains them in a unified way. The model has three important aspects: e-commerce reduces consumers' search costs, involves trade-offs for consumers, and reduces retailing costs. E-commerce reduces consumers' search costs, because on the Web consumers can visit at a low cost virtual shops and learn prices[5], or can use ***shopbots***, software agents that automatically search for price information[6]. E-commerce involves trade-offs for consumers, because buying from a virtual shop does not require a shopping trip, but requires waiting for delivery. E-commerce reduces retailing costs, compared to physical shops, because virtual shops allow savings on ***property costs***, i.e., leases and acquisition of shop and warehouse space, on ***labor costs***, i.e., personnel to attend shops, and on ***inventory costs***, i.e., inventories for showcasing or immediate delivery[7].

In our model, firms decide whether to open virtual shops and set prices, and consumers search for prices. There are two consumer types: new consumers have Internet access, old consumers do not, or do not consider using the Internet an option. New consumers canvass prices through the Web, and then decide if they buy from a virtual or a physical shop. There are two firms: the old firm has a physical shop, the new firm does not. Virtual shops have lower marginal production costs than physical shops.

Since search and waiting for delivery are costly, new consumers accept prices above the minimum charged in the market. This gives firms market power.

The virtual shops' pricing behavior is simple. Virtual shops have the lowest cost and charge the lowest price. Thus, they are not constrained by consumer search, and charge their monopoly price.

---

[4] Other empirical studies are: Bailey (1998), Brynjolfsson & Smith (1999), Chevalier & Goolsbee (2000), Clemons, Hann & Hitt (1999), Ellison & Ellison (2001), Friberg, Ganslandt & Sandstrom (2000), Iyer & Pazgal (2001), Karen, Krishnan, Wolff, & Fernandes (1999), Morton, Zettelmeyer & Risso (2000)).

[5] *Yahoo Internet Life*, August 19 1999, reports that it took 32 minutes to find a hotel in New York using "the old way", while only 6 minutes using "the net way".

[6] E.g., ClickTheButton, DealPilot, www.previewtravel.com for airfares, and www.microsurf.com for mortgages.

[7] On the Web, a banking transaction costs $.1, compared with $.27 at an ATM or $.52 over the phone, and processing an airline ticket costs $1, compared with $8 through a travel agent (*The Economist*, June 26, 1999). USA retailers with no physical presence in a state do not collect local sales taxes, 6%.

The physical shop's pricing behavior depends of whether the old firm has a virtual shop, and on whether the new firm is in the market. Because new consumers have access to lower cost shops, and if waiting for delivery is not too costly, they only accept buying from a physical shop for a lower price than old consumers. When only the new firm opens a virtual shop, if the physical shop charges a lower price acceptable to both consumer types, it earns a lower per consumer profit, if it charges a higher price acceptable only to old consumers, it earns a higher per consumer profit. Thus, the physical shop trades-off ***volume of sales*** and ***per consumer profit***; sometimes it chooses to sell to all consumers, and other times only to old consumers. When both firms open virtual shops, the old firm faces an additional effect, besides the volume of sales and per consumer profit effects. If its physical shop charges a lower price acceptable to both consumer types, half of the new consumers it sells to would otherwise buy from the old firm's virtual shop, where per consumer profit is higher. This causes the old firm to have its physical shop charge a lower price to attract new consumers, only if the virtual shops' cost reduction is small; otherwise it prefers to sell to new consumers only from its virtual shop. We argue that these price equilibria are different from others in search theory, where firms face consumers with different reservation prices.

If the new and old firms' virtual shops have different costs, there will be price dispersion on-line

Since information goods are more convenient to buy on-line, physical shops must charge lower prices than virtual shops to be able to sell them to new consumers.

The firms' incentives to open virtual shops depend on the virtual shops' cost reduction, and the new consumers' reservation price. If cost reduction is small, the new firm has more incentives to open a virtual shop; if cost reduction is large, and the new consumers' reservation price is high, this is no longer true. In fact, the old firm can choose to open a virtual shop when the new firm does not.

The model has two novel features. First, it captures some of the consumers' and firms' trade-offs regarding e-commerce. Second, the production and the search cost distributions are endogenous.

Section 2 presents the basic model, where reservation prices are exogenous, and section 3 characterizes its equilibria. Section 4 discusses the firms' incentives to open virtual shops. Section 5 allows the new and old firm to operate the new technology at different costs. Section 6 presents the model with endogenous reservation prices.

Section 7 discusses price equilibria for information goods. Section 8 discusses related literature. Proofs are in the Appendix.

# 2 The Basic Model

In this section we formalize the firms' opening of a virtual shop and pricing decisions, given consumers' reservation prices, as a 2 stage game. Later we will insert this *Basic* model in a larger game that includes a third stage, where reservation prices are determined.

### (a) The Setting

Consider a retail market for a homogeneous search good that opens for 1 period.

There are 2 alternative *retailing technologies*[8]: a *New*, virtual shop based technology, and an *Old*, physical shop based technology. A *Virtual Shop* has a Web site, where consumers can observe prices and buy, and its logistics is based on the Web. A *Physical Shop* has a physical location, where consumers can observe prices and buy, and its logistics is based on the physical world. A physical shop may have a Web site, but only to post prices[9]. A firm is *Old* if it has a physical shop, opened before the game, and *New* if it does not.

The game has 2 stages. In stage 1 firms choose whether to open virtual shops. In stage 2 firms choose prices. Then consumers buy, delivery takes place, agents receive their payoffs, and the market closes.

Subscript *j* refers to firms and we index a new and an old firm by: $n, o$. Subscripts *t* refers to shops and we index a new firm's virtual shop, an old firm's virtual shop, and a physical shop by: $vn, vo, p$.

### (b) Consumers

There is a unit measure continuum of risk neutral consumers of 2 types. *New* consumers, a proportion $l \in (0,1]$, have Internet access; *Old* consumers do not. At price $p$ a consumer demands $D(p)$, where $D(.)$ is a differentiable, decreasing, bounded function, with a bounded inverse.

Consumers ignore the prices of individual shops, and can only learn them by visiting the shops. Old consumers visit the physical shop's physical location, and if offered a price no higher than $r$, where $D(r) \equiv 0$, buy

---

[8] Technologies that make products available for use or consumption. This concept is related to that of a *distribution channel* (see Kotler (1994)).

and receive the product. When there are no virtual shops, new consumers behave similarly. Otherwise new consumers canvass prices through the Web[10]. They have the list of Web sites, obtained, e.g., from a search engine, but do not know to which type of shop the directions correspond. At the end of section 6 we explain the role of the assumptions that consumers do not know beforehand to which type of shops the Web sites correspond, that the physical shop has a Web site, and that when there are virtual shops, new consumers canvass prices through the Web. We assume that:

**(H.1)** Each new consumers picks randomly which Web site to visit, from the set he has not sampled yet.

The new consumers' reservation price for a type $t$ shop is $r_t$. When new consumers visit a new (old) firm's virtual shop, if offered a price no higher than $r_{vn}$ ($r_{vo}$), they buy, and wait for delivery; when they visit a physical shop's Web site, if offered a price no higher than $r_p$, they go to the shop's physical location, buy, and receive the product; otherwise they reject the offer and search again[11]. Visiting a Web site or a physical shop's physical location, and waiting for delivery of the product bought from a virtual shop, involve costs which we will ignore until section 6.

## (c) Firms

There are 2 risk neutral firms: a new and an old firm. If the new firm decides not to open a virtual shop, it exits the game with a $0$ payoff. Opening a virtual shop involves a set-up cost, $K \in (0, +\infty)$. The probability with which firm $j$ opens a virtual shop is $a_j$; let $a = (a_n, a_o)$. At the end of stage 1 $a$ is observed by all players. If at least 1 virtual shop opens, the physical shop creates its own Web site, where it posts its price.

Marginal production costs are constant for both shop types. The marginal cost of shop $t$ is $c_t$. A virtual shop has a lower marginal cost than a physical shop. Let $c_p \in (0, r)$ and $c_{vn} = c_{vo} = c_v = c_p - D_c$, where $c_p$ is the common production cost, and $D_c \hat{I}(0, c_p]$ is the ***production cost reduction*** induced by the new technology. All players know $(c_p, c_v)$.

---

[9] Bailey (1998) and Brynjolfsson & Smith (1999) found that, e.g., Cody's and Powell's Books, posted prices on the Web, but only sold at their physical locations.

[10] In 2000, about half of the US car buyers will use the Internet. Most of them not to buy, but to obtain information to bargain lower prices out of local dealers (*The Economist*, February 14, 1998).

[11] As an alternative to sequential search new consumers could use shopbots. Shopbots give consumers a sample of between 20 to 40 prices at a low fixed search cost. Thus, although they to not give consumers perfect information, or necessarily identify the lowest price, contrary to popular belief, shopbots can be approximated by a ***newspaper search technology*** (Braverman (1980), Salop & Stiglitz (1977), Wilde & Schwartz (1979)), i.e., perfect information at a fixed cost. The firms' trade-offs, between selling to only to high search cost consumers, or selling to high and to low search cost consumers, or between using only one retail technologies, or using both, remain qualatatively the same in this alternative setting, except that price equilibrium would be in mixed strategies.

The old firm can charge different prices at its 2 shops[12]. Shop $t$'s price and per consumer profit are $p_t$ and $p(p_t; c_t) := (p_t - c_t)D(p_t)$. Let $\hat{p}_t := \arg\max_p p(p; c_t)$. Assume that $p(.)$ is strictly quasi-concave in $p$, and that even for the maximum cost reduction the physical shop can charge $\hat{p}_v$ without losses, i.e., $c_p < \hat{p}_v$ for $D_c = c_p$. Shop $t$'s expected consumer share and expected profit are: $f_t(p_t)$ and $P(p_t; c_t) := p(p_t; c_t) f_t(p_t)$. The new and old firm's net expected profits are: $V^n := [P(p_{vn}; c_v) - K]a_n$ and $V^o := P(p_p; c_p) + [P(p_{vo}; c_v) - K]a_o$.

A firm's stage 1 ***strategy***, is a rule that for every firm type, says with which probability a firm should open a virtual shop. A firm's stage 2 ***strategy***, is a rule that for each history and shop type, says which price a shop should charge. A firm's ***payoff*** is expected profit, net of the investment expenditure.

**(d) Equilibrium**

A subgame perfect Nash ***Equilibrium*** is: an opening and a pricing rule, for each shop and firm type, $\{(a_j^*, p_t^*) j = n, o; t = vn, vo, p\}$, such that:

**(E.1)** Given any $r_t$ and $a$, firms choose $p_t^*$ to solve problems: $\max_{p_{vn}} V^n$ and $\max_{\{p_{vo}, p_p\}} V^o$;

**(E.2)** Given any $r_t$, and $p_t^*$, firms choose $a_j^*$ to solve problem: $\max_{a_j} V^j$.

# 3 Equilibrium of the Basic Model

In this section we construct the basic model's equilibrium by working backwards. First, given reservation prices and the profile of opening of virtual shops decisions, we derive the firms' equilibrium prices. Virtual shops charge their monopoly price. The physical shop charges sometimes the new consumers' reservation price, other times its monopoly price. Second, given reservation prices and equilibrium prices, we derive the firms' equilibrium opening of virtual shop's rule. Either firm sometimes opens a virtual shop, sometimes does not. There are 6 types of equilibria, depending on whether firms choose to open a virtual shop, and whether the physical shop sells to all or only to old consumers.

---

[12] Barnesandnoble.com charges different prices than Barnes and Noble's physical shops.

## 3.1 Stage 2: The Price Game

In this sub-section we characterize equilibrium prices.

The number of shops that charge a price acceptable to new consumers, i.e., $p_t \leq r_t$, $t = vn, vo, p$, is $a$. If virtual shop $t$ charges a price higher than $r_t$, it makes no sales; if it charges a price no higher than $r_t$, given **(H.1)** and that there is a continuum of new consumers, its expected consumer share is $l/a$. Thus, for $0 < a$:

$$\phi_t(p; \rho_t) = \begin{cases} 0 & \Leftarrow \rho_t < p \\ \lambda/\alpha & \Leftarrow p \leq \rho_t \end{cases} \quad t = vn, vo$$

(we omit $a$ and $l$ in $f_t$).

If the physical shop charges a price higher than $r$, it makes no sales; if it charges a price higher than $r_p$, but no higher than $r$, it sells to old consumers, $1 - l$; if it charges a price no higher than the $r_p$, its expected consumer share is $l/a + 1 - l$. Thus, for $0 < a$:

$$\phi_p(p; \rho_p) = \begin{cases} 0 & \Leftarrow r < p \\ 1 - \lambda & \Leftarrow \rho_p < p \leq r \\ \lambda/\alpha + 1 - \lambda & \Leftarrow p \leq \rho_p \end{cases}$$

(we omit $a$, $l$ and $r$ in $f_p$).

We assume that $r_p < \hat{p}_p$, which rules out the uninteresting cases, where although virtual shops exist, the physical shop is able to sell to new consumers at $\hat{p}_p$, its monopoly price. We assume also that $r_t$ is strictly higher than the lowest of the prices consumers can find if they search:

**(H.2)** $\min\{p_{t'}\} < \rho_t \qquad t' \neq t$

This assumption rules out equilibria which are not subgame perfect in the larger model, if search and waiting for delivery are costly. It follows that costly search and impatience give firms market power[13], since they lead new consumers to accept prices above the minimum charged in the market. By **(H.2)**, $0 < a$.

When neither firm opens a virtual shop, $a = (0, 0)$, the industry is a monopoly. The number of shops that charge a price acceptable to new consumers when firms play $(a_n, a_o)$ in stage 1 is $a^{a_n a_o}$; $a^{00} = 1$.

Next we examine the case where only the new firm opens a virtual shop, and hence the industry's supply side consists of the physical shop, and the new firm's virtual shop. The value of $r_p$ for which the old firm is indifferent between charging $p_p = r_p$, and charging $p_p = \hat{p}_p$, given $a = (1, 0)$ and $p_{vn} \leq r_{vn}$, is $p_o^s$, i.e. $p(p_o^s(1); c_p)[1/2 + 1 - 1] \equiv p(\hat{p}_p; c_p)(1 - 1)$. We assume that when the old firm is indifferent between selling to both consumers types and selling only to old consumers, it chooses the latter.

**Proposition 1:** If $a = (1, 0)$, then: **(i)** $p_{vn}^* = \hat{p}_v$; **(ii)**

$$p_p^* = \begin{cases} \rho_p & \Leftarrow p_o^s(\lambda) < \rho_p \\ \hat{p}_p & \Leftarrow \rho_p \leq p_o^s(\lambda) \end{cases}$$

where $p_o^s(.)$ is decreasing, and $p_o^s(1) = c_p$. §

Since the new firm's virtual shop charges the lowest price in the market, and given **(H.2)**, it is never constrained by consumer search and always charges $\hat{p}_v$. The physical shop also benefits from the market power generated by costly search, and from being the only shop old consumers can buy from, by charging a higher price than the new firm's virtual shop. However, it is constrained by consumer search, if it is beneficial to sell to both consumer types[14]. Let $\hat{l}(r_p) := (p_o^s)^{-1}(r_p)$. If $r_p$ is high, i.e., $p_o^s(1) < r_p$, or alternatively, if $l$ is large, i.e., $\hat{l}(r_p) < l$, the old firm wants to sell to both consumers types, so reduces its price below $\hat{p}_p$ and charges $r_p$ (figure 1). If $r_p$ is low, i.e., $r_p < p_o^s(1)$, or alternatively, if $l$ is small, i.e., $l < \hat{l}(r_p)$, the old firm wants to sell only to old consumers and charges $\hat{p}_p$.[15] The higher is $l$, the more willing is the physical shop to lower its price to sell to new consumers. From **Proposition 1**:

$$\alpha^{10} = \begin{cases} 2 & \Leftarrow \; p_o^s(\lambda) < \rho_p \\ 1 & \Leftarrow \; \rho_p \leq p_o^s(\lambda) \end{cases}$$

---

[13] The ability to raise price above marginal cost.
[14] And the threat of a second search by new consumers is credible, i.e., $r_p < \hat{p}_p$.

When the old firm does not open a virtual shop and charges $r_p$ instead of $\hat{p}_p$, it sells to $l/2$ new consumers, earning an additional $p(r_p; c_p)(l/2)$, the *Volume of Sales* effect, but loses $-[p(\hat{p}_p; c_p) - p(r_p; c_p)]$ per old consumer, and a total of $-[p(\hat{p}_p; c_p) - p(r_p; c_p)](1-l)$, the *per Consumer Profit* effect. Thus, the physical shop trades-off volume of sales and per consumer profit[16].

**[Insert figure 1 here]**

When the physical shop charges $r_p$, new consumers search only once[17]; otherwise new consumers may search twice, until they find the virtual shop.

When only the new firm opens a virtual shop there can be 2 types of price equilibria. In both the virtual shop charges $\hat{p}_v$. The physical shop at a *Competing* equilibrium charges $r_p$, and at a *Segmentation* equilibrium charges $\hat{p}_p$. The *Competing* equilibrium occurs when $(r_p, l)$ are large, and the *Segmentation* equilibrium occurs when $(r_p, l)$ are small.

Next we examine the case where both firms open virtual shops, and hence the industry's supply side consists of a physical shop and 2 virtual shops. The level of $r_p$ for which the old firm is indifferent between its physical shop selling to both consumer types, and selling only to old consumers, given $a = (1, 1)$ and $p_t \leq r_t$, $t = vn, vo$, is $p_m^s$, i.e., $p(\hat{p}_v; c_v)(l/3) + p(p_m^s(1, D_c); c_p)[l/3 + 1 - l] \equiv p(\hat{p}_v; c_v)(l/2) + p(\hat{p}_p; c_p)(1-l)$. We assume that when the old firm is indifferent between its physical shop charging $p_p = r_p$, and charging $p_p = \hat{p}_p$, it chooses the latter; and that for $D_c = c_p$, $2 < p(\hat{p}_v; c_v)/p(\hat{p}_p; c_p)$, which can be interpreted as the *Large Cost Reduction Opportunities* case. The value of $D_c$ for which $p(\hat{p}_v; c_v)/p(\hat{p}_p; c_p) \equiv 2$, is $D_c^c$.[18]

---

[15] When $r_p < p_p^*$ the physical shop could shut its Web site.

[16] The physical shop could price discriminate between new and old consumers, by, e.g., offering coupons at its Web site. It might, however, be reluctant to do so, because when informed about them, old consumers could perceive these price differences as unfair. See Sinha (2000) for a discussion of this issue.

[17] The option to search serves only as a credible, out of equilibrium threat, constraining the old firm's price decisions.

**Proposition 2:** If $a = (1, 1)$, then: **(i)** $p_{vn}^* = p_{vo}^* = \hat{p}_v$; **(ii)**

$$p_p^* = \begin{cases} \hat{p}_p & \text{for } \Delta_c \in [\Delta_c^c, c_p] \\ \begin{cases} \rho_p & \Leftarrow p_m^s(\lambda, \Delta_c) < \rho_p \\ \hat{p}_p & \Leftarrow \rho_p \leq p_m^s(\lambda, \Delta_c) \end{cases} & \text{for } \Delta_c \in (0, \Delta_c^c) \end{cases}$$

where $p_m^s(.)$ is decreasing in $l$, increasing in $D_c$, $p_o^s(1) < p_m^s(1, D_c)$, and $p_m^s(1, D_c) \in (c_p, \hat{p}_p)$. §

Note that $p_{vn}^* = p_{vo}^* = \hat{p}_v$ is an expression of Diamond's (1971) paradox[19]. From **Proposition 2**:

$$\alpha^{11} = \begin{cases} 2 & \text{for } \Delta_c \in [\Delta_c^c, c_p] \\ \begin{cases} 3 & \Leftarrow p_m^s(\lambda, \Delta_c) < \rho_p \\ 2 & \Leftarrow \rho_p \leq p_m^s(\lambda, \Delta_c) \end{cases} & \text{for } \Delta_c \in (0, \Delta_c^c) \end{cases}$$

When the old firm opens a virtual shop, it faces another *per Consumer Profit* effect, now with respect to new consumers, besides the *Volume of Sales* effect, $p(r_p; c_p)(l/6)$, and the previous *per Consumer Profit* effect with respect to old consumers, $-[p(\hat{p}_p; c_p) - p(r_p; c_p)](1 - l)$. If its physical shop charges $r_p$ instead of $\hat{p}_p$, half of the new consumers it sells to, $l/6$, would otherwise buy from its own virtual shop, causing a loss of $-[p(\hat{p}_v; c_v) - p(r_p; c_p)](l/6)$. This additional effect causes the old firm to only want to reduce its physical shop's price below $\hat{p}_p$ to attract new consumers, if cost reduction is small, i.e., $D_c \leq D_c^c$. Otherwise, the old firm prefers to sell to new consumers only from its virtual shop. And when the old firm's physical shop does reduce its price to attract new consumers, it does so for higher reservation price values than when it does not open a virtual shop, $p_o^s < p_m^s$ (figure 2). Even when all consumers have Internet access, $l = 1$, if cost reduction is small, and $p_m^s < r_p$, the old firm still sells from the physical shop, $p_m^s(1, D_c) < \hat{p}_p$, since this allows it to have a new consumer share of $2l/3$ instead of $l/2$.

---

[18] Price $\hat{p}_v$ depends on $c_v$, and thus on $D_c$. If $2 < p(\hat{p}_v; c_v)/p(\hat{p}_p; c_p)$, $D_c^c \leq c_p$, whereas if $p(\hat{p}_v; c_v)/p(\hat{p}_p; c_p) \leq 2$, $D_c^c = c_p$.

12**[Insert figure 2 here]**

When both firms open virtual shops there is a ***Competing*** and a ***Segmentation*** equilibrium. A ***Competing*** equilibrium, exists when $D_c$ is small and $(r_p, 1)$ are large, and a ***Segmentation*** equilibrium exists when either $D_c$ takes intermediate values and $(r_p, 1)$ are small, or when $D_c$ is large.

The price equilibria of case $a = (1, 1)$ are different from other search theory equilibria where firms must choose whether to sell only to high reservation price consumers, or to sell also to low reservation price consumers (e.g., Braverman (1980), Burdett & Judd (1983), Rob (1985), Salop & Stiglitz (1977), Varian (89), Wilde & Schwartz (1979)), because the old firm's problem is not just whether to sell to low reservation price consumers, but also how to sell to them, since it can do so either through its virtual or its physical shop.

Next we examine the case where only the old firm opens a virtual shop, and hence the industry's supply side consists of the old firm's physical and virtual shops.

**Proposition 3:** If $a = (0, 1)$, then: **(i)** $p^*_{vo} = \widehat{p}_v$ ; **(ii)** $p^*_p = \widehat{p}_p$.    §

Now since the old firm is alone in the industry, it has no incentive to reduce its physical shop's price below $\widehat{p}_p$. Any new consumer its physical shop might attract is stolen from its virtual shop, where per consumer profit is no smaller. And, if all consumers have Internet access, $1 = 1$, since $r_p < \widehat{p}_p$, the physical shop has zero sales, which could be interpreted as ***Shutting Down***. From **Proposition 3**: $a^{01} = 1$.

When only the old firm opens a virtual shop there is a ***Segmentation*** equilibrium.

Table 1 summarizes the price equilibria's main features.

**[Insert table 1 here]**

---

[19] Low cost shops charge their monopoly price, regardless of how low the search cost is, and how many shops there are. See Davis & Holt (1996) for experimental evidence.

Next we order the equilibrium price distributions for the various stage 1 decision profiles. The market equilibrium price distribution when in stage 1 firms play $(a_n, a_o)$ is $F^{a_n a_o}(.)$; $F'(.) \prec F(.)$ means "distribution $F(.)$ dominates distribution $F'(.)$ in the first-order stochastic sense".[20]

**Proposition 4:** (i) $F^{10} \prec F^{01} \prec F^{00}$; (ii) $F^{11} \prec F^{01}$ §

Since $F^{00}$ dominates all other distributions, if at least 1 virtual shop opens, prices fall due to 2 effects. The *Cost Reduction* effect, is the fall on prices caused by the production cost reduction induced by e-commerce. The *Price Competition* effect, is the fall on prices induced by the physical shop lowering its price to compete for new consumers with the new firm's virtual shop. If the new firm opens a virtual shop prices fall, since $F^{10} \prec F^{00}$ and $F^{11} \prec F^{01}$. If the old firm opens a virtual shop, prices only fall for sure if it is the only virtual shop, $F^{01} \prec F^{00}$. When the new firm opens a virtual shop, the old firm opening a virtual shop puts more weight on the left tail of the price distribution. But since $p_o^s \leq p_m^s$, the physical shop charges a no lower price than when it has no virtual shop. Thus, $F^{11}$ and $F^{10}$ are not comparable in the first-order stochastic sense.

## 3.2 Stage 1: The Opening of Virtual Shops Game

In this sub-section we characterize the equilibrium opening rule and establish existence of equilibrium.

Firm $j$'s net profit when in stage 1 firms play $(a_n, a_o)$, and after both firms and consumers play optimally is $V_{a_n a_o}^j$. The difference between firm $j$'s net profits when it opens a virtual shop, and when it does not, given that firm $j'$ plays $d = 0, 1$ in stage 1 is $D_{1|d}^j$, e.g., $D_{1|1}^o = V_{11}^o - V_{10}^o$ and $D_{1|1}^n = V_{11}^n - V_{01}^n = V_{11}^n$. Firm $j$'s *Expected Incremental Profit* of opening a virtual shop is $S_j := a_{j'} D_{1|1}^j + (1 - a_{j'}) D_{1|0}^j$, $j' \neq j$.

Firm $j$'s optimal stage 1 decision is to open a virtual shop if its expected incremental profit is positive.

**Proposition 5:** Equilibrium exists. §

---

[20] Distribution $F(.)$ **Dominates** distribution $F'(.)$ in the **First-Order Stochastic** sense if $F(.) \leq F'(.)$, for all $p$.

Given the range of values the $S_j$'s can take, virtually any profile of decisions to open a virtual shop can be an equilibrium. In section 4 we describe equilibrium profiles for particular cases.

# 4 Incentives to Open Virtual Shops

In this section we discuss the firms' incentives to open virtual shops. We show that if cost reduction is small, the new firm opens a virtual shop when the old firm does, and also when the old firm does not; otherwise the old firm may open a virtual shop when the new firm does not.

We start by discussing how opening a virtual shop impacts the new and old firms' profits. Since the new firm has no physical shop, opening a virtual shop enables it to sell to new consumers. The **Business Creating** effect, is the increase in the new firm's profit from opening a virtual shop, $p(\hat{p}_v; c_v)(1/a)$. The impact of opening a virtual shop on the old firm's profit can be decomposed in 3 effects. The old firm can sell to new consumers through its physical shop. But, if it opens a virtual shop it can sell to them at a lower cost. The **Cost Reduction** effect, is the increase in the old firm's profit from selling to new consumers through its virtual shop, instead of its physical shop, $[p(\hat{p}_v; c_v) - p(p_p^*; c_p)](1/m_c)$, where $1/m_c$ is the proportion of new consumers that buy from the old firm's virtual shop, but that would buy from the physical shop if the old firm did not open a virtual shop. By opening a virtual shop when the new firm also does, the old firm improves its ability to sell to new consumers. The **Market Penetration** effect, is the increase in the old firm's profit, due to the rise in its new consumers' share, from opening a virtual shop when the new firm also does, $p(\hat{p}_v; c_v)(1/m_p)$, where $1/m_p$ is the proportion of new consumers that buy from the old firm's virtual shop, but that would buy from the new firm's virtual shop if the old firm did not open a virtual shop. By opening a virtual shop when the new firm also does, the old firm can sell to new consumers through its virtual shop, and have its physical shop sell only to old consumers. The **Price Discrimination** effect, is the increase in the old firm's profit from switching from a *Competing* to a *Segmentation* equilibrium, $[p(\hat{p}_p; c_p) - p(r_p; c_p)](1-1)$.[21]

---

[21] The *Cost Reduction* effect is present in $D_{I_0}^o$, and in $D_{I_1}^o$ if $p_o^s \le r_p$; the *Market Penetration* effect is present in $D_{I_1}^o$ if $r_p < p_o^s$ for $D_c^c < D_c$, and if $r_p \in [p_o^s, p_m^s)$ for $D_c \le D_c^c$; *the Price Discrimination* effect is present in $D_{I_1}^o$ if $p_o^s \le r_p$ for $D_c^c < D_c$, and if $r_p \in [p_o^s, p_m^s)$ for $D_c \le D_c^c$.

The next lemma orders the firms incremental payoffs. The value of $D_c$ for which $p(\hat{p}_v;c_v)/p(\hat{p}_p;c_p) \equiv 3/2$, is $\hat{D}_c(3/2)$.

**Lemma 1: (i)** If $D_c \in (0,\hat{D}_c(3/2))$, then $D^o_{1|0} < V^n_{11}$, and thus $\max\{D^o_{1|0}, D^o_{1|1}\} < V^n_{11} \le V^n_{10}$. **(ii)** If $p^s_o(1) < r_p$ and $D_c \in (D^c_c, c_p]$, then $V^n_{11} = V^n_{10} < D^o_{1|0}$, and thus $D^o_{1|1} < V^n_{10} < D^o_{1|0}$. **(iii)** $\hat{D}_c(3/2) < D^c_c$.  §

Since the new firm's consumer share is no bigger when the old firm opens a virtual shop than when it does not: $V^n_{11} \le V^n_{10}$. When both firms open a virtual shop, the **Business Creating** effect dominates the **Market Penetration**, **Cost Reduction** and **Price Discrimination** effects: $D^o_{1|1} \le V^n_{11}$, and thus $D^o_{1|1} \le V^n_{11} \le V^n_{10}$.

Expression $V^n_{1d} - D^o_{1|0}$, $d = 0, 1$, equals the difference between the **Business Creating** and the **Cost Reduction** effects. If $D_c$ is small, i.e., $D_c < \hat{D}_c(3/2)$, the **Business Creating** effect dominates the **Cost Reduction** effect, $D^o_{1|0} < V^n_{11}$, and thus: $\max\{D^o_{1|0}, D^o_{1|1}\} < V^n_{11} \le V^n_{10}$. If, however, $D_c$ is large, i.e., $D^c_c < D_c$, and the physical shop competes for new consumers, i.e., $p^s_o(1) < r_p$ or alternatively $\hat{I}(r_p) < 1$, the **Cost Reduction** effect dominates the **Business Creating** effect, $V^n_{10} < D^o_{1|0}$. Furthermore, for $a = (1,0)$ the model has a **Competing** equilibrium, and for $a = (1,1)$ a **Segmentation** equilibrium. Thus, $V^n_{11} = V^n_{10}$, and consequently: $D^o_{1|1} < V^n_{11} = V^n_{10} < D^o_{1|0}$.

Next we characterize the opening of virtual shops equilibrium profiles, for **Lemma 1**'s 2 cases. To focus on pure strategy equilibria, we assume that when a firm is indifferent between opening and not opening a virtual shop it chooses the former[22]. We assume also that $0 \le \max\{D^o_{1|0}, V^n_{10}\}$, which rules out $a^* = (0,0)$. Let $w = (1, c_p, D_c, K, r_p)$.

**Proposition 6: (i)** If $D_c \in (0, \hat{D}_c(3/2))$, then:

$$a^* = \begin{cases} (1,0) \Leftarrow \{w \mid \Delta^o_{1|1} < 0\} \\ (1,1) \Leftarrow \{w \mid 0 \le \Delta^o_{1|1}\} \end{cases}$$

---

[22] Although Proposition 6 is not exhaustive, an equilibrium in pure strategies does exist for all parameter values. See Mazón & Pereira (2001a).

**(ii)** If $p_o^s(1) < r_p$ and $D_c \hat{I} \left(D_c^c, c_p\right]$, then:

$$a^* = \begin{cases} (0,1) & \Leftarrow & \{w \mid V_{10}^n < 0 \leq \Delta_{1|0}^o\} \\ (1,0) & \Leftarrow & \{w \mid \Delta_{1|1}^o < 0 \leq V_{10}^n\} \\ (1,1) & \Leftarrow & \{w \mid 0 \leq \Delta_{1|1}^o\} \end{cases}$$

§

If $D_c$ is small, the new firm opens a virtual shop when the old firm does, and also when the old firm does not: $D_{1|1}^o < 0$. We interpret this as the old firm having less incentive to open a virtual shop.

**[Insert figure 3 here]**

For $(1, D_c) \hat{I} \left(0, \hat{I}_c(r_p)\right] \times \left(0, \hat{D}_c(3/2)\right)$, $D_{1|1}^o$ and $V_{10}^n$ are increasing in $(1, D_c)$, and decreasing in $K$ (figure 3 (i)). If $(1, D_c)$ are small, $D_{1|1}^o < 0 \leq V_{10}^n$, and thus $a^* = (1,0)$; for larger values of $(1, D_c)$, $0 \leq D_{1|1}^o$, and thus $a^* = (1,1)$. Although the model is static, this could explain why typically new firms opened virtual shops before old firms. Initially, $(1, D_c)$ were small because few consumers had Internet access, and firms did not fully understand the new technology. Overtime, more consumers gained access to the Internet, and firms learned how to use the new technology.

If however, $(1, D_c)$ are large, the old firm may open a virtual shop when the new firm does not: $V_{10}^n < 0 \leq D_{1|0}^o$.

For $(1, D_c) \hat{I} \left(\hat{I}_c(r_p), 1\right] \times \left(D_c^c, c_p\right]$, $D_{1|0}^o$, $D_{1|1}^o$, and $V_{10}^n$, are increasing in $(1, D_c)$, and decreasing in $K$ (figure 3: (ii)). If $(1, D_c)$ are larger than $(\hat{I}_c(r_p), D_c^c)$ but not by much, $V_{10}^n < 0 \leq D_{1|0}^o$, and thus $a^* = (0,1)$; for larger values of $(1, D_c)$, again $0 \leq D_{1|1}^o$, and thus $a^* = (1,1)$.

The model shows how the old firm might have less incentive to open a virtual shop if cost reduction is small. But, if cost reduction is large, the old firm need not have less incentive. In fact, it can choose to open a virtual shop when the new firm does not.

It has been claimed that an old firm may be reluctant to use e-commerce for fear of its virtual shop, with supposedly a lower per consumer profit, stealing business from its physical shop, i.e., *Self-Cannibalization*[23]. Note first that cannibalization is part of the *Cost Reduction* effect. If a virtual shop has lower costs than a physical shop, and market power, this intra-firm transfer of new consumers is profitable[24]. Second, if an old firm opens a virtual shop it can increase its new consumers' share, *Market Penetration* effect, and price discriminate between new and old consumers, *Price Discrimination* effect, both of which are also profitable[25]. In the next section we add some comments on this issue.

# 5 Firm Asymmetry

In this section, we discuss how a possible asymmetry between the new and old firm with respect to the new technology, affects the firms' pricing behavior and incremental profits.

We assumed that the new and old firm are equally capable of achieving the new technology's cost reduction. However, if virtual shops require new forms of organization that take advantage of the new technology's low cost of information processing and transmission, if integrating virtual and physical shop retailing is hard, and if old firms' employees resist the new technology because it devalues their skills, the new firm might achieve larger retailing cost reductions than the old firm. To model this asymmetry let $c_{vo} = c_p - (1-e)D_c$, where $e \in [0,1]$ measures the *efficiency loss* of the old firm relative to the new firm.

The firms' pricing behavior remains unchanged, except for $a=(1,1)$, where $p_{vo}^* = min\{\hat{r}_{vo}, \hat{p}_{vo}\}$, $\hat{p}_{vo} \in [\hat{p}_v, \hat{p}_p]$. This implies that there may be price dispersion among virtual shops[26].

Now $D_{1|0}^o < V_{11}^n$, if $D_c$ is small, or, if $D_c$ is large but $e$ is also large (figure 4). And $V_{10}^n < D_{1|0}^o$, if $D_c$ is large and $e$ is small. And for $e$ large enough $D_{1|0}^o < V_{11}^n$ for all values of $D_c$.

**[Insert figure 4 here]**

---

[23] Toys"R"Us invested $80 million to launch a virtual division, but Robert Mogg, the man in charge, resigned, claiming that the firm was afraid of competing with its own physical shops (*El País*, September 5, 1999). Alba et al. (1997): "E-commerce offers an advantage to retailers that have low penetration (…). On the other hand, companies with high penetration might experience significant cannibalization of its in-shop sales, making e-commerce less attractive".

[24] Baseball Express, claims that its Web site stole sales from its catalogue, but that selling on the Web is more profitable (*The New York Times*, September, 2, 1999).

[25] Ward & Morganosky (2000) and Ward (1999) also arrive to a negative conclusion on cannibalization, although for different reasons.

The *Cost Reduction* effect becomes $\left[p\left(min\{r_{vo}, \hat{p}_{vo}\}; c_{vo}\right) - p\left(p_p^*; c_p\right)\right]\left(1/m_c\right)$, which can be negative if $r_{vo}$ is small and $c_{vo}$ large, giving some justification to the fear of *Self-Cannibalization*. The *Market Penetration* effect $p\left(min\{r_{vo}, \hat{p}_{vo}\}; c_{vo}\right)\left(1/m_p\right)$ also becomes smaller.

# 6 Endogenous Reservation Prices

In this section we add to the model, a third stage where reservation prices are determined, given consumers' search and waiting costs. The game consists of 3 stages. The first 2 unfold as in the basic model. In stage 3 consumers make their search and purchase decisions; then delivery takes place, agents receive their payoffs, and the market closes.

To complete the model we introduce the following costs. Visiting the physical shop involves cost, $s \in (0, +\infty)$, which includes the opportunity cost of the time spent, and associated expenses like driving. Visiting a Web site involves cost, $s - \Delta_s$, which includes the opportunity cost of the time spent, and associated expenses like phone calls and Internet fees, and where, $\Delta_s \in (0, s)$, is the *search cost reduction* induced by the new technology. Waiting for delivery of a product bought from a virtual shop involves cost, $d$, that results from deferring consumption. Searching Web sites is instantaneous, a consumer may observe any number of prices, and may at any time accept any offer received to date. Let $S(p) := \int_p^\infty D(t)dt$. The surplus of a consumer that buys from a type $t$ shop at price $p_t$, is $S(p_t) - u_t$, where $u_t = s$ if $t = p$, and $u_t = d$ if $t = vn, vo$. Let $\Delta S := S(\hat{p}_v) - S(\hat{p}_p)$.

Old consumers, and new consumers when $\alpha = (0,0)$, visit the physical shop, and if offered a price $p \leq r$ buy and receive the product, getting a surplus of $S(p) - s$. When $\alpha \neq (0,0)$, new consumers first visit a Web site chosen at random: **(H.1)**.[27] Then, they decide if they accept the best offer at hand and terminate search; or if they reject it, retaining the option to recall it later, and visit one of the other shop's Web sites. If new consumers have visited all shops, they accept the offer with the highest surplus.

---

[26] See Baye & Morgan (2000) and Ayer & Pazgal (2000) for alternative ways of generating price dispersion on-line.
[27] This first step is usually absent in the search literature since it is assumed that consumers get their first price observation for free.

A consumer's *information set* just after his *k*-th search step consists of all previously observed prices. A consumer's stage 3 *strategy*, $s$, is a stopping rule, that for any sequence of observations, says if search should stop or continue. A consumer's *payoff* is the expected consumer surplus, net of the search expenditure.

A subgame perfect Nash *Equilibrium* is: a stopping rule for new consumers, and an opening and a pricing rule, for each shop and firm type, $\{(a_j^*, p_t^*, s^*)\ j = n, o; t = vn, vo, p\}$, such that:

**(E.0)** Given any $a$ and $p_t$, new consumers choose $s^*$ to maximize net expected surplus;

**(E.1)** Given any $a$, and $s^*$, firms choose $p_t^*$ to solve problems: $max_{p_{vn}} V^n$ and $max_{\{p_{vo}, p_p\}} V^o$;

**(E.2)** Given $s^*$ and $p_t^*$, firms choose $a_j^*$ to solve problem: $max_{a_j} V^j$.

Next we characterize the new consumers equilibrium search behavior for $a \neq (0,0)$.

When $a = (1,0)\ (0,1)$, new consumers' search may involve 3 steps. In step 3, consumers know both prices, and the optimal strategy is to accept $p_t$, if $S(p_t) - u_t \geq S(p_{t'}) - u_{t'}$. In step 2, a consumer who was offered $p_t$ at the shop he choose to visit at random in step 1, gains $[S(p_{t'}) - u_{t'} - S(p_t) + u_t]$ by searching. Search is optimal if and only if $s - D_s < [S(p_{t'}) - u_{t'} - S(p_t) + u_t]$. Let $r_t^{a_n a_o}$ equate the marginal search cost, $s - D_s$, to the marginal benefit, when firms play $(a_n, a_o)$ in stages 1:

$$[S(p_{t'}) - u_{t'} - S(\rho_t^{a_n a_o}) + u_t] = \sigma - \Delta_\sigma \qquad t' \neq t \tag{1}$$

The new consumers' optimal search rule is to accept offer $p_t$ and terminate search, if $p_t \leq r_t^{a_n a_o}$, and reject offer $p_t$ and proceed to step 3, if $p_t > r_t^{a_n a_o}$.[28] Equation (1) defines implicitly reservation price function, $r_t^{a_n a_o} = R_t^{a_n a_o}(.)$, which is increasing in $p_{t'}$. Also, $R_t^{a_n a_o}(p_p, s, D_s, d)$, $t = vn, vo$, are increasing in $s$, and decreasing in $D_s$ and $d$, and $R_p^{a_n a_o}(p_t, D_s, d)$, $t = vn, vo$, is increasing in $d$, and decreasing in $D_s$. It is straightforward to show that the maximum price for which new consumers accept the physical shop's offer in step 3, is strictly smaller than $r_p^{10}$. Thus the physical shop cannot charge a price higher than $r_p^{10}$, expecting that it will be rejected in step 2, but accepted in step 3.

---

[28] See Reinganum (1979) or Benabou (1993).

If $p_{vn}^* = p_{vo}^*$, $r_{vn}^{10} = r_{vo}^{01}$.

The next lemma establishes the parameter restrictions required for $r_p^{a\,n^a\,o} < \hat{p}_p$ and **(H.2)**.

**Lemma 2: (i)** If $D_s < d$, then $p_t < r_p^{a\,n^a\,o}$, $t = vn, vo$. **(ii)** If $d < 2s - D_s$, then $p_p < r_t^{a\,n^a\,o}$, $t = vn, vo$. **(iii)** If $d < DS + D_s$, then $r_p^{a\,n^a\,o} < \hat{p}_p$. §

From **Lemma 2: (i)-(ii)**, **(H.2)** follows from search and waiting for delivery being costly. Since new consumers have access to lower cost shops, and from **Lemma 2: (iii)**, if waiting for delivery is not too costly, new consumers only accept buying from the physical shop for a lower price than old consumers: $r_p < \hat{p}_p < r$. From now on let $d \in (D_s, D_s + min\{DS, 2(s - D_s)\})$.

When $a = (1,1)$, new consumers' search may consist of 4 steps. Steps 3 and 4 are similar to steps 2 and 3 of the previous 2 cases. In step 2, there are 2 Web sites to sample. Let $S(p_{t'}) - u_{t'} < S(p_{t''}) - u_{t''}$. If $S(p_t) - u_t < S(p_{t''}) - u_{t''}$, a new consumer who is offered $p_t$, gains $[S(p_{t'}) - u_{t'} - S(p_t) + u_t]$ by sampling shop $t'$, and gains $[S(p_{t''}) - u_{t''} - S(p_t) + u_t]$ by sampling shop $t''$. If $S(p_{t''}) - u_{t''} \leq S(p_t) - u_t < S(p_{t'}) - u_{t'}$, new consumers' expect to gain $[S(p_{t'}) - u_{t'} - S(p_t) + u_t]/2$ by searching, since they reject shop $t''$'s offer. The optimal search rule is to hold reservation price $r_t^{11}$ which equates the marginal benefit to the marginal cost:

$$\begin{cases} \frac{1}{2}[S(p_{t'}) - u_{t'} - S(\rho_t^{11}) + u_t] + \frac{1}{2}[S(p_{t''}) - u_{t''} - S(\rho_t^{11}) + u_t] = \sigma - \Delta_\sigma & \Leftarrow S(r_t^{11}) - u_t < S(p_{t''}) - u_{t''} \\ \frac{1}{2}[S(p_{t'}) - u_{t'} - S(\rho_t^{11}) + u_t] = \sigma - \Delta_\sigma & \Leftarrow S(p_{t''}) - u_{t''} \leq S(r_t^{11}) - u_t < S(p_{t'}) - u_{t'} \end{cases} \quad (2)$$

Equation (2) also defines implicitly reservation price function, $R_t^{11}(.)$, which is non-decreasing in $p_{t'}$ and $p_{t''}$. As before, it is straightforward to show that the maximum price for which new consumers accept the physical shop's offer, is smaller in step $t$ than in step $t + 1$, $t = 2, 3$. Lemma 2 holds, and if $s < (S(\hat{p}_p) + D_s)/2$, in equilibrium consumers always have a strictly positive net surplus.

Equilibrium prices are as in section 3 and **Proposition 5** holds.

We conclude this section by explaining the role of the assumptions that consumers do not know beforehand to which type of shops the Web sites correspond, that the physical shop has a Web site, and that when there are virtual shops, new consumers must canvass prices through the Web. In Reinganum (1979) and Benabou (1993), firms have different costs, play pure strategies, and consumers do not know the firms' costs $c$. Thus, although consumers know the firms' equilibrium price rule $p^*(.)$, they do not know the firms' equilibrium prices, $p^*(c)$, and have to search to learn them. In our model consumers know $(c_p, c_v)$. Thus, the first 2 assumptions ensure that meaningful search occurs. Alternatively, consumers could not know $(c_p, c_v)$. This would mean developing a more complicated incomplete information game. If when there are virtual shops, we allow new consumers to choose weather to canvass prices through the Web, the *Competing* Equilibrium becomes non-generic, unless we also make new consumers heterogeneous with respect to, e.g., $d$, which has an expository cost. Finally, although these assumptions are intended to ensure meaningful search in a simple setting, they are not without justification, since as we argue in footnotes 9 and 10, some consumers in some markets behave similarly.

# 7 Information Goods

In this section we discuss the price equilibria for the case where $d = 0$. We argue that for this case, virtual shops' prices may be higher than the physical shop's price.

Restriction, $0 < D_s < d$, potentially rules out information goods, for which $d$ is small, possibly zero. So assume $d = 0$. From (1), $R_p^{a_n a_o}(p_t, s, D_s, d) < p_t$, $t = vn, vo$. **Lemma 2: (ii)** and **(iii)** holds.

Equilibrium prices are as in section 3, with an important difference. When $a = (1,0)$ and $p_o^s < r_p$, or $a = (1,1)$ and $p_m^s < r_p$, if $0 < D_s < d$, $p_t^* = \hat{p}_v < r_p = p_p^*$, whereas if $d = 0$, $p_p^* = r_p < \hat{p}_v = p_t^*$, $t = vn, vo$. If buying from a virtual shop is more convenient than buying from a physical shop, the physical shop must charge a lower price than virtual shops to sell to new consumers.

Case $d = 0$ illustrates an intuitive point. Functionally identical goods sold through different retailing technologies, acquire different attributes. E-commerce reduces prices through the *Price Competition* and *Cost*

*Reduction* effects. If in addition consumers value negatively e-commerce's attributes, relative to those of other retailing technologies, consumers will only buy on-line if compensated by lower prices, which pushes prices further down. If however, consumers value positively e-commerce's attributes, they will pay for the convenience of buying on-line, and the net effect can be such that prices are higher on-line than off-line.

# 8 Related Literature

This section inserts the paper on the literature. Our paper relates to 3 literature branches. First, to the e-commerce marketing literature: Alba, Lynch, Weitz, Janiszewski, Lutz, Sawyer & Wood (1997), Bakos (1997), Lal & Sarvary (1998), Peterson, Balasubramanian, & Bronnenberg, (1997), Zettelmeyer (1997). Bakos (1997) presents a model of circular product differentiation, where consumers search for prices and product characteristics, i.e., locations. All consumers have Internet access. If search costs for price and product information are separated, and if e-commerce lowers the former, prices decrease; if it lowers the latter, prices can increase.

Second, our paper relates to the literature that analyzes competition between alternative retailing technologies: Balasubramanian (1998), Bouckaert (2000), Friberg, Ganslandt & Sandstrom (2000), Michael (1994), and Legros & Stahl (2000). Balasubramanian (1998) and Bouckaert (2000) use a model of circular product differentiation to analyze competition between catalogue and physical shop retailing. Physical shops are located on the circumference, and catalogue firms at the center of the circle. The presence of a catalogue firm lowers prices, and the number of physical shops in the market.

Third, our paper relates to the advertising and markets for information literatures: Baye & Morgan (1998), Caillaud & Jullien (2000), Ellison & Ellison (2001), Iyer & Pazgal (2000), and Kephart & Greenwald (1999). Baye & Morgan (1998) examine the interaction between markets for information and the product market they serve. They show that the product market can exhibits price dispersion even if consumers are fully informed. Kephart & Greenwald (1999) investigate the impact of shopbots on markets. Shopbots allow users to choose the number of searches, and make search cost depend only weakly on the number of searches, i.e., nonlinear, leading to a more extensive search.

In two companion papers (Mazón & Pereira 2001a, 2001b) we .

# References

**Alba**, J., **Lynch**, J., **Weitz**, B., **Janiszewski**, C., **Lutz**, R., **Sawyer**, A. & **Wood**, S., 1997, "Interactive Home Shopping: Consumer, Retailer, and Manufacturer Incentives to Participate in Electronic Marketplaces", *Journal of Marketing*, **61**, 38-53

**Bailey**, J., 1998, "Electronic Commerce: Prices and Consumer Issues for Three Products: Books, Compact Discs, and Software", OCDE/GD(98)4

**Balasubramanian**, S., 1998, "Mail versus Mall: A Strategic Analysis of Competition between Direct Marketers and Conventional Retailers", *Marketing Science*, **17**, 181-95

**Bagwell**, K. & **Ramey**, G., 1994, "Advertising and Coordination", *Review of Economic Studies*, **61**, 153-172

**Bakos**, J., 1997, "Reducing Buyer Search Costs: Implications for Electronic Marketplaces", *Management Science*, **43**, 1676-1692

**Bakos**, Y., **Lucas**, H., **Oh**, W., **Simon**, G., **Viswanathan**, S. & **Weber**, B., 2000, "The Impact of Electronic Commerce on the Retail Brokerage Industry", Stern School

**Baye**, M. & **Morgan**, J., 1998, "Information Gatekeepers and Competitiveness of Homogeneous Product Markets", *American Economic Review* (forthcoming)

**Benabou**, R., 1993, "Search, Market Equilibrium, Bilateral Heterogeneity and Repeat Purchases", *Journal of Economic Theory*, **60**, 140-63

**Bouckaert**, J., 1999, "Monopolistic Competition with a Mail Order Business", *Economics Letters*, **66**, 303-10

**Braverman**, A., 1980, "Consumer Search and Alternative Market Equilibria", *Review of Economic Studies*, **47**, 487-502

**Brown**, J. & **Goolsbee**, 2000, "Does the Internet Make Markets More Competitive? Evidence from the Life Insurance Industry", *University of Chicago*

**Brynjolfsson**, E. & **Smith**, M., 1999, "Frictionless Commerce? A Comparison of Internet and Conventional Retailers", *Sloan School of Management*

**Burdett**, K. & **Judd**, K., 1983, "Equilibrium Price Dispersion", *Econometrica*, **51**, 955-69

**Caillaud**, B. & **Jullien**, B, 2000, "Competing Cybermediaries", *IDEI*

**Clemons**, E., **Hann**, I. & **Hitt**, L., 1999, "The Nature of Competition in Electronic Markets: An Empirical Investigation of Online Travel Agent Offerings", *The Wharton School*

**Davis**, D. & **Holt**, C., 1996, "Consumer Search Costs and Market Performance", *Economic Inquiry*, **34**, 133-151

**Diamond, P.,** 1971, "A Model of Price Adjustment", *Journal of Economic Theory*, **3**, 156-168

**Ellison**, G. & **Ellison**, S., 2001, "Search, Obfuscation, and Price Elasticities on the Internet"; M.I.T.

**Friberg**, R., **Ganslandt**, M. & **Sandstrom**, M., 2000, "E-Commerce and Prices: Theory and Evidence", *Stockholm School of Economics, EFI*

**Iyer**, G. & **Pazgal**, A., 2000, "Internet Shopping Agents: Virtual Co-Location and Competition",

**Kephart**, J. & **Greenwald**, A., 1999, "Shopbot Economics", *IBM Institute for Advanced Commerce*

**Kreps**, D. & **Wilson**, R., 1982, "Sequential Equilibria", *Econometrica*, **50**, 863-894

**Kotler**, P., 1994, *Marketing Management*, 8th ed., Prentice Hall

**Lal**, R. & **Sarvary**, M., 1998 "When and How is the Internet Likely to Decrease Competition?", *Harvard Business School Management Science*

**Legros**, P. & **Stahl**, K., , "", ECARE

**MacMinn, R.,** 1980, "Search and Market Equilibrium", *Journal of Political Economy*, **88**, 308-327

**Mazón**, C. & **Pereira**, P., 2001a, "Who Benefits from Electronic Commerce?", *U.C.M.*

**Mazón**, C. & **Pereira**, P., 2001b, "Patterns of Adoption of Electronic Commerce", *U.C.M.*

**Michael**, S., 1994, "Competition in Organizational Form: Mail Order versus Retail Stores, 1910-1940", *Journal of Economic Behavior and Organization*, **23**, 269-286.

**Morton**, **F**., **Zettelmeyer**, **F**., & **Risso**, **J**., 2000, "Internet Car Retailing", *Haas Business School*

**Peterson**, **R**., **Balasubramanian**, **S**., & **Bronnenberg**, **B**., 1997, "Exploring the Implications of the Internet for Consumer Marketing", *Journal of the Academy of Marketing Science*, **25**, 329-346

**Reinganum**, **J**., 1979, "A Simple Model of Equilibrium Price Dispersion", *Journal of Political Economy*, **87**, 851-858

**Rob**, **R**., 1985, "Equilibrium Price Distributions", *Review of Economic Studies*, **52**, 487-504

**Rosenthal**, **R**., 1980, "A Model in Which an Increase in the Number of Sellers Leads to a Higher Price", *Econometrica*, **48**, 1575-79

**Stahl, D.**, 1989, "Oligopolistic Pricing with Sequential Consumer Search", *American Economic Review*, **79**, 700-12

**Salop**, **S**. & **Stiglitz**, **J**., 1976, "Bargains and Rip-offs: A Model of Monopolistically Competitive Price Dispersion", *Review of Economic Studies*, **44**, 493-510

**Sinha**, **I**., 2000, "Cost Transparency: The Net's Real Threat to Prices and Brands", *Harvard Business Review*, March-April

**Subramani**, **M**. & **Walden**, **E**., 1999, "The Dot Com Effect: The Impact of E-Commerce Announcements on The Market Value of Firms", Carlson School of Management, University of Minnesota

**Varian**, **H.,** 1980, "A Model of Sales", *American Economic Review*, **70**, 651-59

**Ward**, **M**. & **Morganosky**, **M**., 2000, "Does Online Retailing Cannibalize Other Marketing Channels?", *University of Illinois, Urbana-Champaign*

**Ward**, **M**. 1999, "Will E-commerce Compete more with Traditional Retailing or Direct Marketing?", *University of Illinois, Urbana-Champaign*

**Wilde**, **L**. & **Schwartz**, **A**., 1979, "Equilibrium Comparison Shopping", *Review of Economic Studies*, **46**, 543-53

**Zettelmeyer**, **F**., 2000, "Expanding to the Internet: Pricing and Communications Strategies When Firms Compete on Multiple Channels", *Journal of Marketing Research (Forthcoming)*

# Graphical Appendix

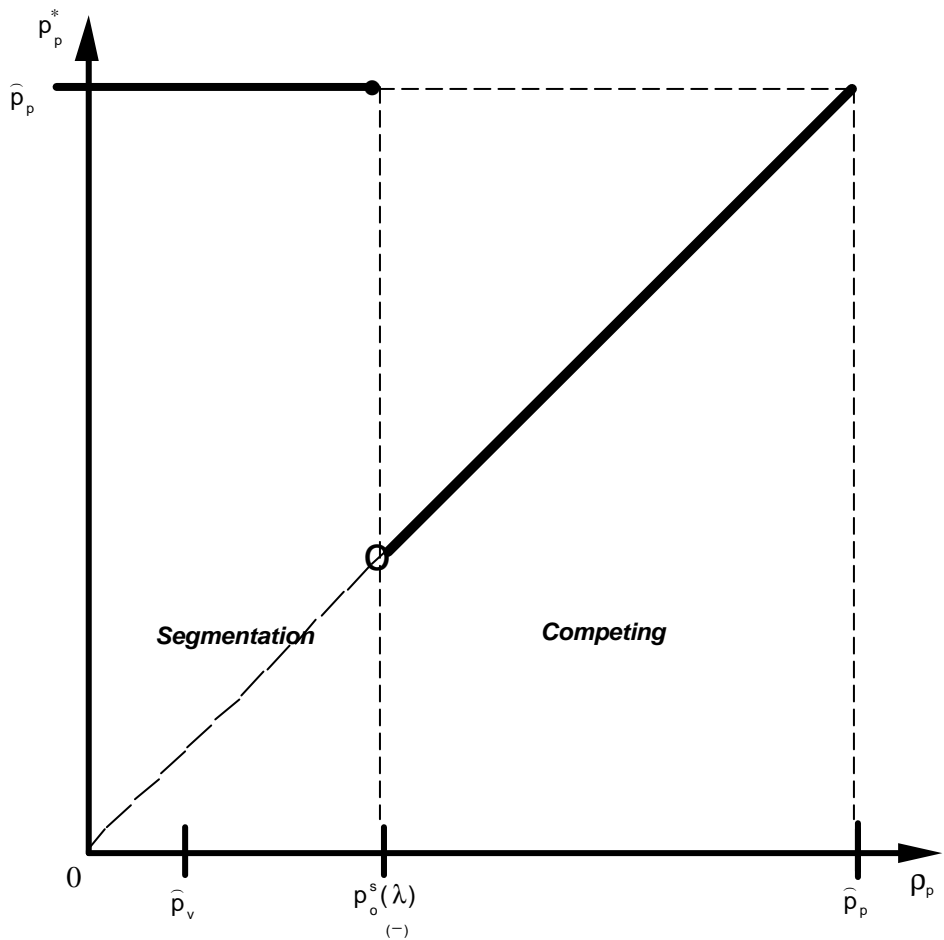

**Figure 1 : Physical Shop's Price**

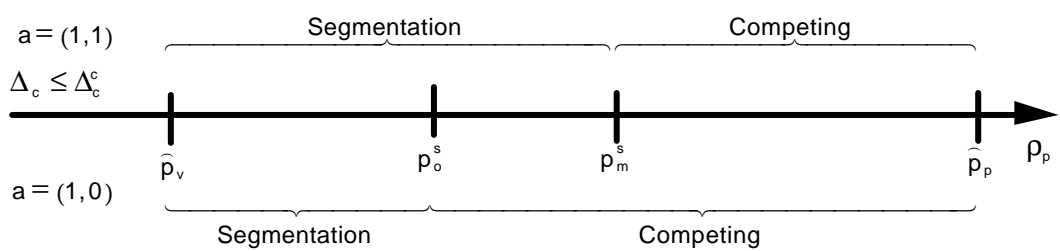

**Figure 2: Price Equilibria**

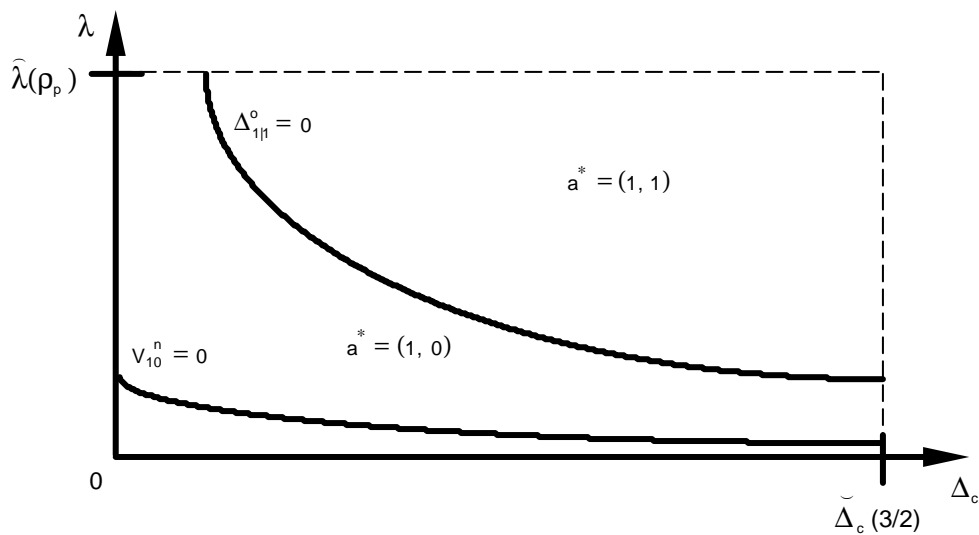

**Figure 3 (i) : Opening of Virtual Shops Equilibrium Profiles**

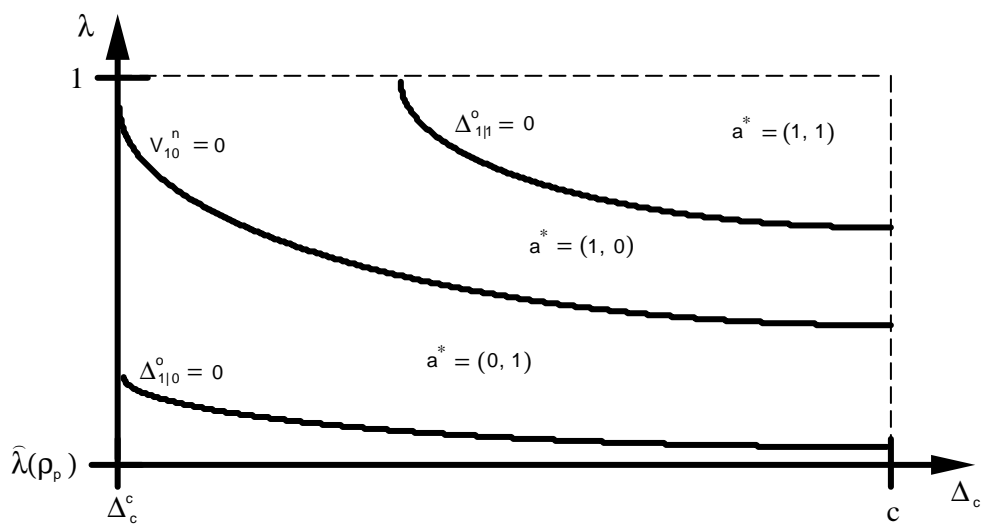

**Figure 3 (ii) : Opening of Virtual Shops Equilibrium Profiles**

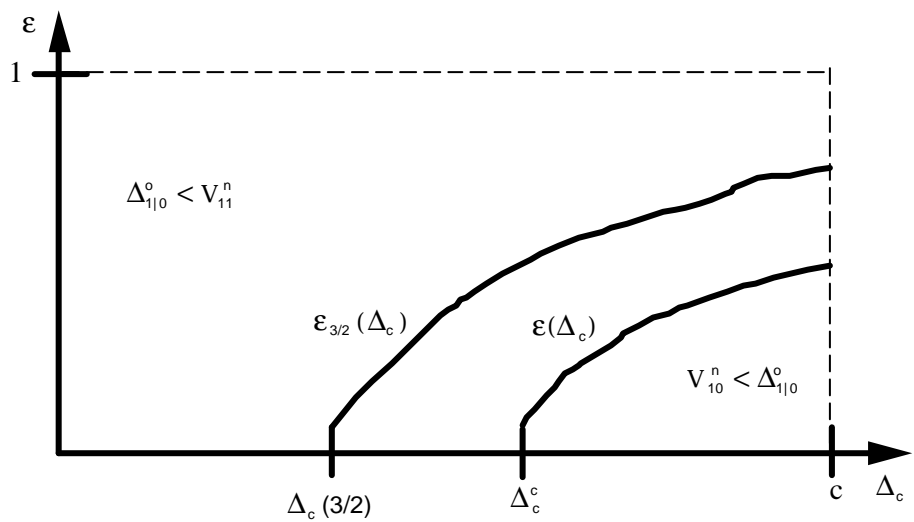

**Figure 4: Cost Asymmetry**

**Table 1: Summary of Model's Price Equilibria**

| a | | | $p_p^*$ | Share vn | Share vo | Share p | Equilibrium |
|---|---|---|---|---|---|---|---|
| $(1,0)$ | $p_o^s < r_p$ | | $r_p$ | $1/2$ | – | $1/2 + 1 - 1$ | Competing |
| | $r_p \leq p_o^s$ | | $\widehat{p}_p$ | $1$ | – | $1 - 1$ | Segmentation |
| $(0,1)$ | – | | $\widehat{p}_p$ | – | $1$ | $1 - 1$ | Segmentation |
| $(1,1)$ | $D_c \leq D_c^c$ | $p_m^s < r_p$ | $r_p$ | $1/3$ | $1/3$ | $1/3 + 1 - 1$ | Competing |
| | | $r_p \leq p_m^s$ | $\widehat{p}_p$ | $1/2$ | $1/2$ | $1 - 1$ | Segmentation |
| | $D_c^c < D_c$ | | $\widehat{p}_p$ | $1/2$ | $1/2$ | $1 - 1$ | Segmentation |